\documentclass[12pt]{iopart}

\usepackage{iopams}
\usepackage{graphicx}
\usepackage{subfigure}

\begin{document}

\title{Anharmonic mixing in a magnetic trap}
\author{R Gommers\dag\ddag, B J Claessens\dag, C J Hawthorn\dag\S, H C W
Beijerinck\dag\ and E J D Vredenbregt\dag}
\address{\dag\ Physics Department, Eindhoven University of Technology, Postal Box 513, Eindhoven, The Netherlands}
\address{\ddag\ Physics and Astronomy Department, University
College London, Gower Street, London WC1E 6BT, United Kingdom}
\address{\S\ School of Physics, University of Melbourne, Parkville, Vic. 3010, Australia}
\ead{r.gommers@ucl.ac.uk}

\begin{abstract}
    We have experimentally observed re-equilibration of a
    magnetically trapped cloud of metastable neon atoms after it
    was put in a non-equilibrium state. Using numerical
    simulations we show that anharmonic mixing, equilibration
    due to the collisionless dynamics of atoms in a magnetic trap, is the dominant
    process in this equilibration. We determine the dependence of its
    time on trap parameters and atom temperature. Furthermore we
    observe in the simulations a resonant energy exchange between the radial and
    axial trap dimensions at a ratio of trap frequencies
    $\omega_r/\omega_z=3/2$. This resonance is explained by a
    simple oscillator model.
\end{abstract}

\section{Introduction}
    Magnetic traps have become a standard and inexpensive tool in
    cold atom physics in recent years~\cite{Esslinger}. Together with the
    application of cooling techniques
    such as evaporative cooling and sympathetic cooling, they have
    enabled the formation of Bose-Einstein condensates~\cite{Anderson} and, more
    recently, degenerate Fermi gases~\cite{Regal}. The most commonly used
    species for cold atom experiments (Rb, Cs, Na) are easily
    cooled to the micro-Kelvin regime, where the atomic dynamics
    in a magnetic trap is usually well described by regular
    harmonic motion. For other species such as the noble gases~\cite{Robert,Tempelaars} and
    the group II elements~\cite{Simien,Daily} however, the lowest attainable
    temperatures can be in the milli-Kelvin regime. In this regime
    the atomic dynamics is more complicated, with coupling between
    the different dimensions of the trap becoming important.

    Anharmonic mixing is a process that couples the motion of
    atoms in different dimensions of a magnetic trap. This coupling
    enables redistribution of energy over the coupled dimensions. Therefore,
    anharmonic mixing has to be considered when applying
    techniques that rely on the motion of atoms in a trap, such as
    evaporative cooling~\cite{Luiten}, Doppler cooling in a magnetic trap~\cite{Helmerson} and
    rethermalization experiments aimed at measuring a scattering
    length~\cite{Monroe}. In order to investigate on what timescale anharmonic
    mixing plays a role, we perform a Monte Carlo simulation of a
    cloud of atoms in a Ioffe-Quadrupole magnetic trap. We determine
    the dependence of this timescale on trap parameters and on atom
    temperature. Surkov
    \textit{et. al.}~\cite{Surkov} have investigated anharmonic
    mixing in the limit of low temperature where it can be treated
    as a perturbation of regular harmonic motion, but to the best
    of our knowledge this is the first paper discussing anharmonic
    mixing at higher temperatures. We will
    also present here an experiment showing that anharmonic mixing
    does play an important role in the atomic dynamics, and can
    even be the dominant mechanism of energy redistribution under
    certain conditions. To conclude we will look at possible
    applications of anharmonic mixing.

\section{Monte Carlo calculations}
    A Ioffe-Quadrupole magnetic trap (MT) is one of the most common
    magnetic traps. Its potential is given by
    \begin{eqnarray}\label{TotalPot}
      U(x,y,z)&=\mu[B_{0}^2+(\alpha^2-B_0\beta)(x^2+y^2)+2B_0\beta
      z^2\nonumber\\&+\frac{1}{4}\beta^2(x^2+y^2)^2
      +\beta^2z^4+2\alpha\beta(x^2-y^2)z]^{1/2}-\mu
      B_0,
    \end{eqnarray}
    where $x$ and $y$ are the radial trap dimensions, $z$ is the axial trap dimension,
    $\mu$ is the magnetic moment of the atom, $\alpha$ the gradient of the magnetic field,
    $\beta$ the curvature of the magnetic field and $B_0$ the magnetic bias
    field. In this expression for the trap potential terms of
    order higher than four have been neglected.
    The MT has trap frequencies $\omega_r$ and
    $\omega_z$ in the limit $3k_BT\ll\mu B_0$, where the trap shape is harmonic:
    \begin{equation}\label{HarmPot}
        U(r,z)=\frac{m}{2}(\omega_r^2r^2+\omega_z^2z^2).
    \end{equation}
    Here $m$ is the atomic mass, $r=(x^2+y^2)^{1/2}$ the radial coordinate,
    $\omega_r=[\frac{\mu}{m}(\alpha^2/B_0-\beta)]^{1/2}$ the radial trap frequency,
    and $\omega_z=(\mu\beta/m)^{1/2}$ the axial trap frequency.
    The higher order terms in (\ref{TotalPot}) are in this limit
    negligible compared to the harmonic terms, the term $2\alpha\beta(x^2-y^2)z$ that
    couples the motion in the axial and radial directions is
    therefore absent from (\ref{HarmPot}).

    The starting point of a simulation is an atom cloud in equilibrium in a
    harmonic trap. The initial positions and velocities of the atoms are
    determined by a Monte Carlo method. For the properties of the atom we use the
    values for metastable neon in the $^3P_2$ state, as this is the atom we
    use in our experiment. Two kinds of clouds have been used, namely
    thermal clouds and clouds in which all the atoms have the
    same energy. The former is used to compare the results of the
    simulations with experiments while the latter is used to gain
    a better understanding of the dynamics of atoms at a certain
    energy. During the simulation the position and velocity of each
    atom is determined as a function of time by integrating
    the equations of motion. At the start of the simulation
    the magnetic bias field $B_0$ is adiabatically ramped down,
    thereby increasing the energy of the atoms in the radial
    direction and changing the effective shape of the potential from
    harmonic to almost linear. The ramping is adiabatic if the condition
    $\frac{1}{\omega_r^2}\frac{d\omega_r}{dt}<1$ is fulfilled~\cite{Surkov}.
    The simulation yields the kinetic and potential energies of the
    cloud in every direction as a
    function of time. From the kinetic energies in
    different directions the transfer of energy from one direction
    to the other can be determined, and a comparison with the
    experiment can be made.

\begin{figure}
    \centering{
       \includegraphics[width=0.5\textwidth]{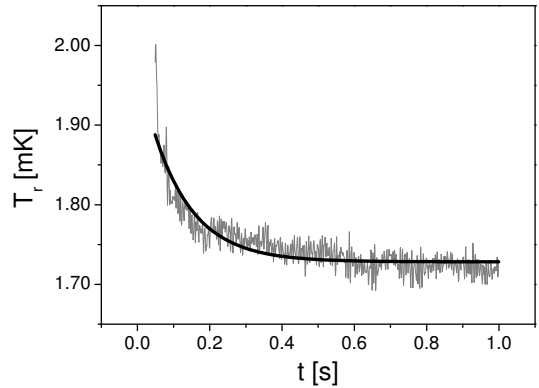}}
     \caption{\textit{Monte Carlo calculation of the radial temperature as a function of time
     for a thermal cloud with an initial temperature of $1$ mK and $10^4$
     atoms (gray). The exponential fit is also shown (black); the temperature during the
     ramping of the bias field ($t=0-50$ ms) is left out.}\label{Thermal}}
    \end{figure}

    We choose the trap parameters at the start of the simulation as
    $\alpha=1\cdot10^4$ G/m, $\beta=47.5\cdot10^4$ G/m$^2$
    and $B_0=99.6$ G. The bias field is then ramped down to 1.5 G.
    These parameters are chosen because they are easily accessible
    in our experiment. A typical calculated result of the temperature
    evolution, or the average kinetic energy,
    in the radial direction of a thermal cloud of 1 mK
    after ramping the bias field is shown in
    Figure~\ref{Thermal}. The first few tens of milliseconds show
    a linear decrease of temperature, after that the decrease is
    approximately exponential with a mixing time of 112 ms for
    this particular cloud with a temperature of 1 mK and a relative
    change in temperature of 8\%.
    The linear decrease of temperature in the first tens of ms is caused
    by high-energy atoms that mix on a timescale of $\frac{1}{2}\omega_z^{-1}=11$ ms.
    This is the average time an atom needs to reach a singular
    point in the potential~\cite{Surkov}, where almost instantaneous mixing can
    occur.

    The dependence of the behavior of a thermal cloud on initial trap
    parameters and cloud temperature is relatively weak. When
    $\alpha$ is varied over the range $0.9-1.2\cdot10^4$ G/m and $B_{0,end}$
    over the range $1.5-20$ G, the mixing time stays
    within the range $100-130$ ms. Only when $\beta$ is decreased,
    a significant change in mixing time occurs:
    $\tau_{mix}\approx400$ ms when $\beta=37.5\cdot10^4$ G/m$^2$.
    The reason for this weak dependence on trap parameters is
    that a significant part of the energy of the cloud is carried by
    atoms that are in a regime where mixing does not occur or
    where it occurs on a timescale of $\frac{1}{2}\omega_z^{-1}$.

\begin{figure}
    \centering{
       \includegraphics[width=0.5\textwidth]{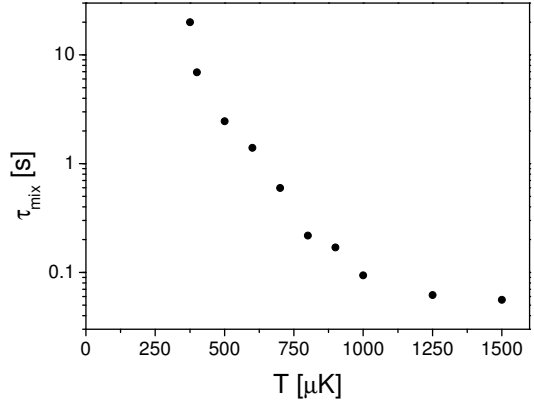}}
     \caption{\textit{Calculated mixing time for atoms with a fixed energy
     $E_{atom}$ drawn as a function of $E_{atom}/3k_B$.}\label{TauTemp}}
    \end{figure}

    To show that indeed the timescale of energy transfer between dimensions of an atom depends
    strongly on the energy of that atom, we now take clouds
    of atoms that all have the \textit{same energy} $E_{atom}$ in
    order to determine the mixing timescale as a function of
    energy. This gives us the average behavior of an atom with
    a certain energy. The result is shown in Figure~\ref{TauTemp}.
    Below $E_{atom}/3k_B=375$ $\mu$K no mixing occurs on the longest
    timescale we considered, i.e. the lifetime of the metastable state
    Ne$^{\ast}(^3P_2)$ of 14.7 s~\cite{Zinner}. For temperatures above 1.5 mK the mixing
    time is determined by $\frac{1}{2}\omega_z^{-1}$ and becomes
    independent of energy.

\begin{figure}
       \begin{center}
         \mbox{
           \subfigure[]{\scalebox{0.5}{\label{TauAlfa}\includegraphics[width=\textwidth]{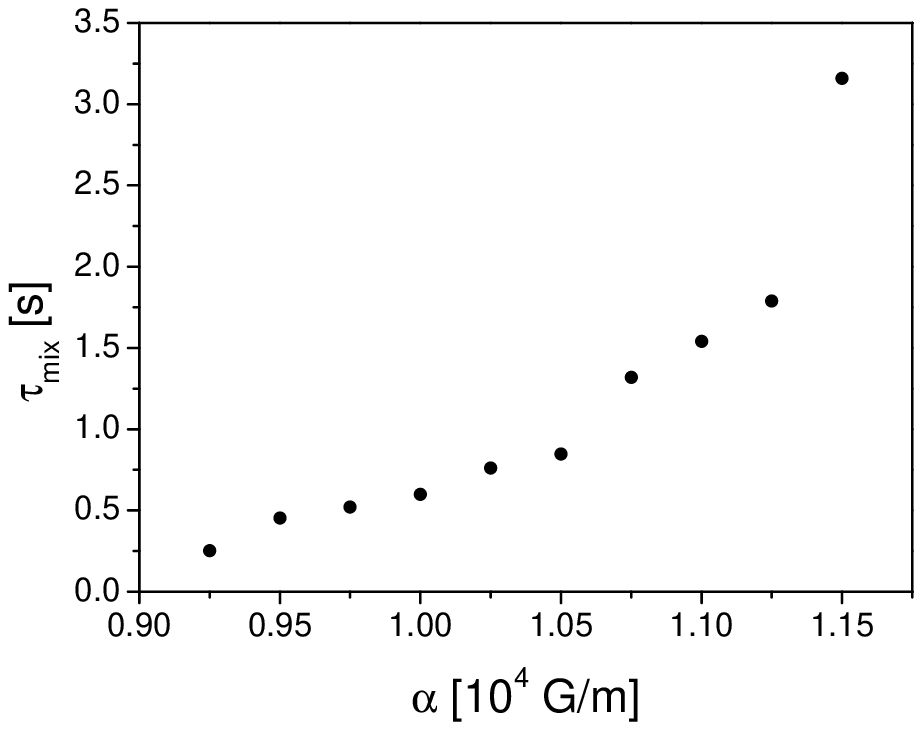}}} \quad
           \subfigure[]{\scalebox{0.5}{\label{TauBeta}\includegraphics[width=\textwidth]{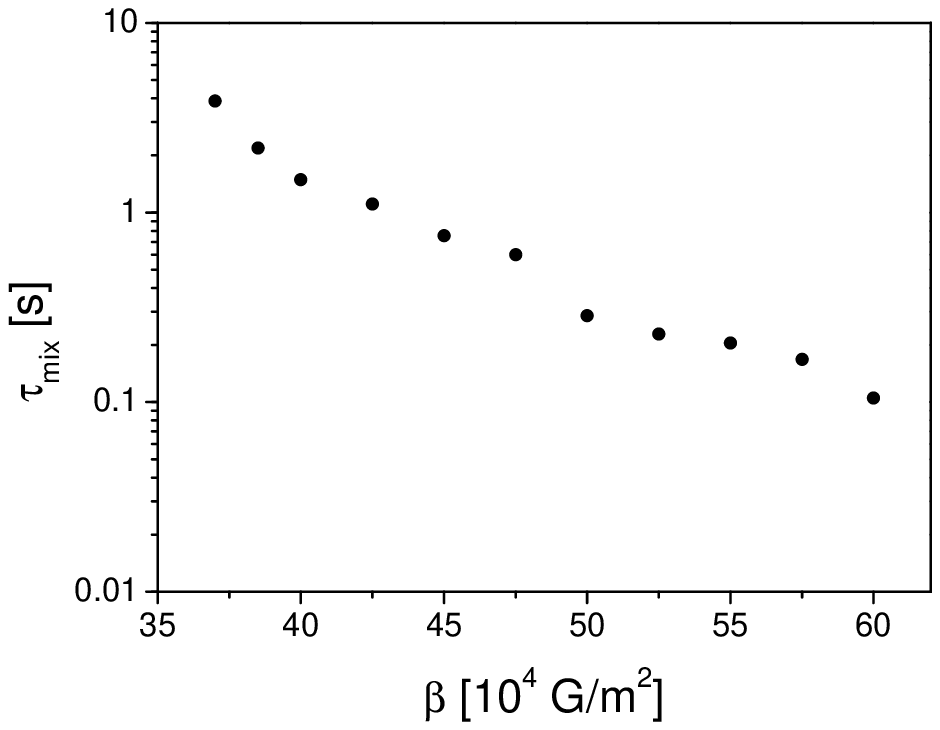}}}
              }
           \caption{\textit{a) Calculated mixing time as a function of magnetic
           gradient in radial direction $\alpha$.
           b) Mixing time as a function of magnetic curvature
           in axial direction $\beta$. Atomic energy
           is $E_{atom}/3k_B=700$ $\mu$K.}\label{TauAlfaBeta}}
        \end{center}
     \end{figure}

    Next, we investigate the dependence of $\tau_{mix}$ on the trap
    parameters $\alpha$ and $\beta$ while keeping the energy
    constant at $E_{atom}/3k_B=700$ $\mu$K. Figure~\ref{TauAlfaBeta} shows that
    $\tau_{mix}$ increases with $\alpha$ and
    decreases with $\beta$. This behavior can be
    understood qualitatively by comparing the strength of the
    harmonic $(\sim\alpha^2-\beta)$ and coupling $(\sim\alpha\beta)$
    terms in (\ref{TotalPot}).

\begin{figure}
    \centering{
       \includegraphics[width=0.5\textwidth]{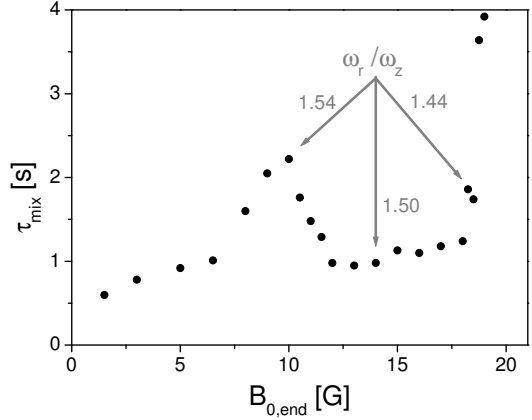}}
     \caption{\textit{Calculated mixing time as a function of bias field after ramping. Atomic energy
     is $E_{atom}/3k_B=700$ $\mu$K.}\label{TauBnul}}
    \end{figure}

    Furthermore we determined the mixing time as a function of the
    value of the bias field at the end of the ramp. When the final
    value of the bias field is higher, less energy is added to the
    atoms and the final temperature of the cloud is lower.
    Therefore we expect that the mixing time becomes
    larger for higher final values of the bias field. As can be seen
    from the result in Figure~\ref{TauBnul} however, there is a
    broad resonance (meaning small mixing time) in the mixing
    time around 14 G. This occurs
    exactly at a ratio of radial and axial trap frequencies of
    $\omega_r/\omega_z=3/2$. The resonance is so broad because the
    radial trap frequency changes only very slowly with bias
    field. The reason for this slow change is that at higher bias
    field the final energy of the cloud is lower, resulting in a higher
    oscillation frequency in the MT because the trap is not harmonic,
    but at the same time the trap is less tightly confining, resulting
    in a lower oscillation frequency. These two effects almost completely
    cancel.

\section{Resonance model}
\begin{figure}
    \centering{
       \includegraphics[width=0.75\textwidth]{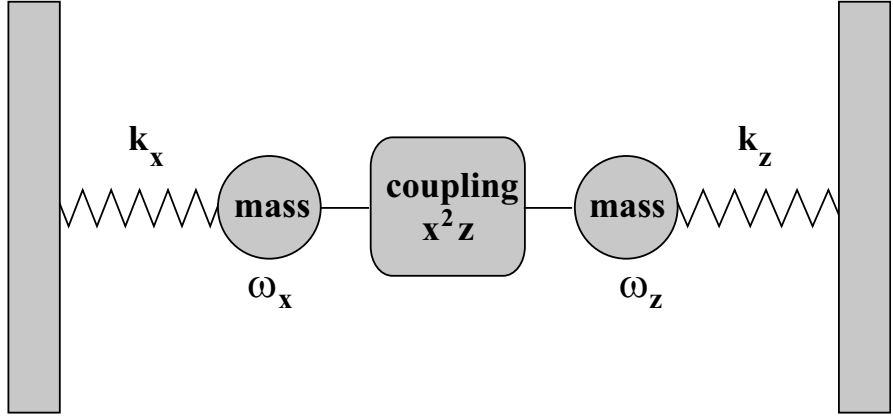}}
     \caption{\textit{Schematic picture of the coupled oscillator
     model. Two one-dimensional oscillators with frequencies
     $\omega_x$ and $\omega_z$ can exchange energy through a
     coupling.}\label{Coupling}}
    \end{figure}

    In order to explain this resonance the dynamics of an atom in the MT is now
    described as two coupled one-dimensional harmonic oscillators,
    whose energies represent the radial and axial energies of the
    atom. The coupling term is chosen to be $W=x^2z$, as
    shown in Figure~\ref{Coupling}. Now we examine the forces
    $F_z$ and $F_x$ that the two oscillators exert on one another, assuming they
    oscillate as $x\sim cos(\omega_xt)$ and $z\sim cos(\omega_zt)$:
    \begin{eqnarray}\label{Force}
        \fl F_z=-\frac{\partial}{\partial z}W=x^2\sim\cos^2(\omega_xt)\sim\cos(2\omega_xt)+1,\nonumber\\
        \fl F_x=-\frac{\partial}{\partial x}W=xz\sim\cos(\omega_xt)\cos(\omega_zt)\sim\cos((\omega_x+\omega_z)t)+\cos((\omega_x-\omega_z)t).
    \end{eqnarray}

    A resonance occurs if an oscillator is driven by a periodic
    force with its eigenfrequency, $F_i\sim cos(\omega_it)$. From
    (\ref{Force}) it can be seen that this happens if
    $\omega_z=2\omega_x$. In general, a coupling term $x^az^b$
    yields a resonance at $a\omega_x=b\omega_z$. When
    potential (\ref{TotalPot}) is linearized a term $x^2z^3$
    appears, explaining the resonance in Figure~\ref{TauBnul}. A
    more formal treatment explaining the resonance is possible by
    transforming the Hamiltonian of an atom in the potential to
    Birkhoff-Gustavson normal form~\cite{Tuwankotta,Quispel},
    but that is beyond the scope of this paper.

\section{Experiment}
    To check our simulation results, we now compare with experiment.
    The only difference between simulation and experiment is that in
    the experiment the atoms can also collide, which we did not
    include in the simulation.
    Therefore it is necessary to consider both anharmonic mixing and collisional
    equilibration of energy. Our setup has been described previously in
    ~\cite{Tempelaars,Kuppens}. Briefly, a discharge source is used to
    create metastable $^3P_2$ neon atoms, and after passing through
    several laser-cooling sections the atomic beam flux is
    $6\cdot10^{10}$ s$^{-1}$~\cite{Tempelaars}. From this atomic beam
    about $10^9$ atoms are loaded into a magneto-optical trap (MOT) at a temperature of
    1 mK. These atoms can be transferred to a Ioffe-Quadrupole magnetic
    trap with almost unity efficiency. To match the size and temperature
    of our MOT the trap parameters are chosen
    as $\alpha=1\cdot10^4$ G/m, $\beta=47.5\cdot10^4$ G/m$^2$
    and $B_0=99.6$ G. The MT has trap frequencies of $\omega_r=(2\pi)376$ Hz and
    $\omega_z=(2\pi)44.7$ Hz.

\begin{figure}
    \centering{
      \includegraphics[width=0.5\textwidth]{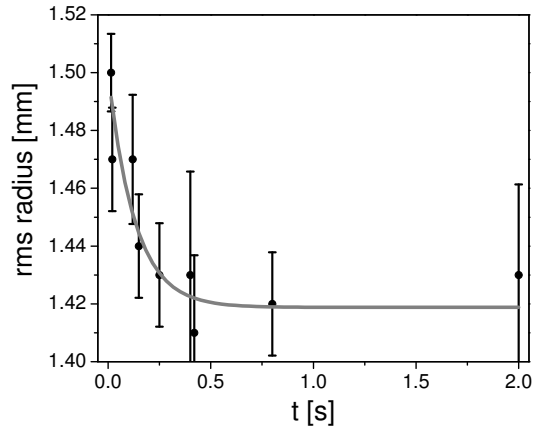}}
     \caption{\textit{Radial rms radius of the atom cloud as a function of
     time. The initial temperature in the MT is $1.0\pm0.1$ mK.}\label{SqueezeExp}}
    \end{figure}

    For this experiment the MOT was operated at a detuning
    of $-2\Gamma$ and at an intensity of 0.5 mW/cm$^2$ for the three
    MOT beams together. After turning off the MOT, a $50$ $\mu$s long $\sigma^+$ spin
    polarization pulse of 2 mW/cm$^2$ was given to put all the atoms in the
    $|m_J=+2>$ state, and then the MT was turned on.
    The atom cloud now has a temperature of approximately 1 mK
    and contains approximately $1\cdot10^9$ atoms. After turning on the MT
    the atoms are held there long enough to allow the atom cloud to reach
    its equilibrium state. Then the bias field is ramped from
    99.6 G to 1.5 G, adding energy to the atoms in the radial direction and
    compressing them spatially. The
    cloud radius in the radial direction as a function of time after the ramping is
    determined by absorption imaging. The radius is directly proportional to
    the potential energy and therefore the temperature because the trap shape
    is linear. A typical result of a series of
    measurements is shown in Figure~\ref{SqueezeExp}.

    The exponential fit gives a characteristic
    equilibration time of $130\pm25$ ms and a relative change in
    width of $5\pm3\%$. This result agrees with the timescale of
    the simulation as found from the data in Figure~\ref{Thermal}.
    If the equilibration would be due solely to
    collisions, this would give a collisional cross-section of
    $\sigma=8\pi\cdot$(200a$_0$)$^2$, with $a_0$ the Bohr radius.
    However, the cross-section at 1 mK can not be larger than the
    unitary limit $\sigma=8\pi(\lambda_{dB}/2\pi)^2=8\pi\cdot$(80a$_0$)$^2$.
    Here $\lambda_{dB}$ is the thermal de Broglie wavelength of the
    atom. Also, we did not observe a
    dependence of the equilibration time on the atomic density, indicating that
    anharmonic mixing is the dominant mechanism here.

\section{Applications}
    An interesting application for anharmonic mixing is to
    use it to enhance the efficiency of Doppler cooling in a magnetic trap. This
    cooling technique is one-dimensional because only along the
    axial direction of an MT all atoms are polarized the same way.
    Cooling in the radial directions can be achieved by
    reabsorption of scattered photons~\cite{Pfau} if the cloud is optically dense.
    If that is not the case anharmonic mixing can cool the radial
    directions, as shown in an early experiment by Helmerson
    \textit{et. al.}~\cite{Helmerson}. By making the mixing time as
    short as possible the cooling can be fast and the atom losses
    as a result of the cooling can be limited.

    Another technique that could benefit from anharmonic mixing is
    evaporative cooling. One of the problems that can occur when
    an atom cloud is cooled evaporatively is gravitational sag,
    in which gravity shifts the equipotential surfaces of atoms in a trap in such a way
    that they do not coincide with the surfaces of constant
    magnetic field. This causes the cooling process to become one
    or two-dimensional, reducing its efficiency~\cite{Thomas}. If the dimensions
    are coupled by anharmonic mixing the evaporation remains
    three-dimensional even when gravity plays a role.

    A situation where anharmonic mixing is undesirable and needs to be
    suppressed is in rethermalization experiments to measure the
    scattering length~\cite{Monroe,Schmidt}. The scattering length is
    one of the properties of an
    atom that determine the feasibility of evaporative cooling and
    reaching the transition point of Bose-Einstein condensation for a given number
    of atoms at a certain density and temperature~\cite{Luiten}.
    It can be measured by observing collisional equilibration
    after ramping the bias field as we did in our simulations and
    experiment. We can conclude that this method will only yield reliable
    results if the timescale on which anharmonic mixing occurs is
    long compared to the collision time.

\section{Conclusion}
    We determined the timescale on which anharmonic
    mixing occurs and the dependence of that timescale on trap parameters
    and atom temperature. We observed a resonance in the mixing time as
    a function of the end bias field, and explained this with a simple
    oscillator model. We verified experimentally that anharmonic mixing does
    indeed occur, and that its timescale can be short compared to
    the timescale needed for collisional equilibration. The
    application of anharmonic mixing to improve Doppler cooling
    and lower-dimensional evaporative cooling is possible.

\end{document}